\documentclass[aps,prb,reprint,superscriptaddress,nobibnotes]{revtex4-1}

\usepackage{siunitx}
\usepackage{graphicx}
\usepackage{hyperref}
\usepackage{amsmath}
\usepackage{bm}

\usepackage{xr}
\begin{document}


\title{First-principles investigation of magnetocrystalline anisotropy 
oscillations in Co\textsubscript{2}FeAl/Ta heterostructures}


\author{Junfeng~Qiao}
\affiliation{Fert Beijing Institute, BDBC, Beihang University, 
Beijing 100191, China}
\affiliation{School of Electronic and Information Engineering, Beihang 
University, Beijing 100191, China}

\author{Shouzhong~Peng}
\affiliation{Fert Beijing Institute, BDBC, Beihang University, 
Beijing 100191, China}
\affiliation{School of Electronic and Information Engineering, Beihang 
University, Beijing 100191, China}

\author{Youguang~Zhang}
\affiliation{Fert Beijing Institute, BDBC, Beihang University, 
Beijing 100191, China}
\affiliation{School of Electronic and Information Engineering, Beihang 
University, Beijing 100191, China}

\author{Hongxin~Yang}
\affiliation{Key Laboratory of Magnetic Materials and Devices, Ningbo 
Institute of Materials Technology and Engineering, Chinese Academy of 
Sciences, Ningbo 315201, China}

\author{Weisheng~Zhao}
\email{weisheng.zhao@buaa.edu.cn}
\affiliation{Fert Beijing Institute, BDBC, Beihang University, 
Beijing 100191, China}
\affiliation{School of Electronic and Information Engineering, Beihang 
University, Beijing 100191, China}


\date{\today}

\begin{abstract}
We report first-principles investigations of magnetocrystalline 
anisotropy energy (MCAE) oscillations as a function of capping layer 
thickness in Heusler alloy Co\textsubscript{2}FeAl/Ta heterostructures. 
Substantial oscillation is observed in FeAl-interface structure. 
According to $k$-space and band-decomposed charge density analyses, 
this oscillation is mainly attributed to the Fermi-energy-vicinal 
quantum well states (QWS) which are confined between 
Co\textsubscript{2}FeAl/Ta interface and Ta/vacuum surface. The smaller 
oscillation magnitude in the Co-interface structure can be explained by the smooth 
potential transition at the interface. These findings clarify that 
MCAE in Co\textsubscript{2}FeAl/Ta is not a local property of the interface 
and that the quantum well effect plays a dominant role in MCAE oscillations 
of the heterostructures. This work presents the possibility of tuning MCAE 
by QWS in capping layers, and paves the way for artificially 
controlling magnetic anisotropy energy in magnetic tunnel junctions.
\end{abstract}


\maketitle

\section{introduction}
With the increasing demand for high-speed and low-power-consumption 
storage devices, intensive researches have been made on 
spin-transfer-torque magnetic random access memory (STT-MRAM). 
The core structure of MRAM is magnetic tunnel junction (MTJ),
\cite{Ikeda2010} 
which is composed of an insulating barrier sandwiched by two 
ferromagnetic (FM) electrodes. The relative orientation of two FM 
electrodes represents two states and can be utilized to store one bit 
information. To realize high storage density, 
the manufacturing process is scaling down to nanometer regime. 
However, the increasing process variations in the fabrication pose serious challenges  
to fundamental physics,\cite{7069229} especially magnetocrystalline anisotropy 
energy (MCAE), which is critical for the thermal stability of the 
relative magnetization orientation of two FM electrodes. 
Previous work reported that to achieve a retention time of 
\num{10} years, an interfacial perpendicular magnetic anisotropy (PMA) 
of \SI[per-mode=symbol]{4.7}{\milli\joule\per\square\meter} is required for device 
sizes scaling down to \SI{10}{\nano\meter}.\cite{Peng2017} However, 
at present the most widely used FM electrode, CoFeB, can commonly 
reach an interfacial PMA of \SI[per-mode=symbol]{1.3}{\milli\joule\per\square\meter} 
when interfaced with MgO tunneling barrier.
\cite{Ikeda2010,PhysRevB.84.054401,PhysRevB.88.184423,RevModPhys.89.025008} 
At the same time tunneling magnetoresistance (TMR) can reach a value of 
\SI{120}{\percent} in CoFeB/MgO/CoFeB MTJ,\cite{Ikeda2010} which needs 
to be improved as well. 

To further promote the development of STT-MRAM, other FM materials are 
under investigation. Heusler alloys
are a big family of ternary intermetallic compounds with 
nearly \num{1500} members.\cite{GRAF20111} 
According to their chemical composition, Heusler alloys can be separated 
into two classes, full Heusler with chemical composition
X\textsubscript{2}YZ ($L2_1$ structure) and half Heusler XYZ ($C1_b$ 
structure), in which X and Y are transition metals, and Z is main 
group element.\cite{0022-3727-43-19-193001} By virtue 
of the broad choices of elements and stoichiometry, many Heusler 
compounds exhibit interesting properties, such as 
half-metallicity,
\cite{PhysRevB.66.174429} 
various Hall effect,
\cite{PhysRevB.73.172417,PhysRevB.80.092408,Chadov2010} 
thermoelectric effect, 
\cite{PhysRevB.72.054116}
topological effect
\cite{Chadov2010} 
and superconductivity, 
\cite{PhysRevB.82.060505} etc.
Among Heusler alloys, Co\textsubscript{2}FeAl (CFA) has attracted lots of 
attention due to its high spin polarization and low magnetic damping 
constant.
\cite{doi:10.1142/S201032471230006X,doi:10.1142/S2010324714400232} 
TMR ratio can reach up to \SI{700}{\percent} at \SI{10}{\kelvin} and 
\SI{330}{\percent} at room temperature (RT) 
in Co\textsubscript{2}FeAl/MgO/CoFe MTJ.
\cite{doi:10.1063/1.3258069} 
Magnetic damping constant, $\alpha$, can reach as low as \num{0.001},
\cite{doi:10.1063/1.3067607} which 
is beneficial for reducing STT switching current. 
Another merit of CFA is its fine lattice matching with MgO. As a result,  
epitaxial growth of CFA(001)[110]$\parallel$MgO(001)[100] can be 
achieved in experiment.\cite{PhysRevB.84.134413} 
All these advantages make CFA a promising candidate for MTJ 
electrode material. Regarding magnetic anisotropy energy (MAE) of CFA, 
experimental and theoretical 
results confirmed that the Co\textsubscript{2}FeAl/MgO interface can 
reach around \SI[per-mode=symbol]{1}{\milli\joule\per\square\meter}.
\cite{doi:10.1063/1.3600645,ADMA:ADMA201401959,PhysRevB.94.104418}
However, as discussed above, 
MAE needs to be optimized further.
Besides, it is crucial to find out effective ways to artificially 
control MAE.

Recently, experimental and theoretical results 
showed that heavy metals (HM) can induce large variations of 
physical properties including MAE when interfaced with FM materials.
\cite{Ouazi2012,PhysRevLett.99.177207,
doi:10.1063/1.4866965,doi:10.1063/1.4972030,doi:10.1063/1.4973477,
Gambardella1130,PhysRevLett.102.257203,doi:10.1063/1.4976517} 
In practical MTJs, a buffer layer at the bottom and a capping layer 
on the top are necessary to improve and protect the FM/MgO/FM core 
structure. Consequently, the choice 
of capping layer provides us a unique way to control MAE of the 
whole structure. On the other hand, when the thickness of these multilayers 
reaches down to atomic scale, quantum mechanical (QM) effects start to 
dominate. One of the most well-known QM effect is quantum well (QW), 
in which the wave functions of the quantum particle are confined by 
potential barriers and the energy levels are quantized. 
In spintronics, the milestone effect,  
giant magnetoresistance (GMR), and its closely related phenomenon, 
interlayer exchange coupling (IEC), are deeply related to QW.  
These effects have been successfully explained by quantum 
interferences due to reflections at the spacer boundaries.
\cite{PhysRevLett.67.1602}
In terms of the influence of quantum well states (QWS) on MAE, early 
theoretical works, using tight-binding formalism and a perturbation 
treatment to spin orbit coupling (SOC), reported the oscillation of 
MAE with respect to Pd layer thickness in Co/Pd system.
\cite{PhysRevB.55.3636} While there also exists other work which supports 
interfacial-MAE in Pd/Co/Pd(111) structure.\cite{PhysRevLett.91.197206}
Other than HM Pd, MAE oscillations with respect to both Co and Cu were 
found in Co/Cu system.\cite{PhysRevB.56.14036} 
Since the IEC effects are prominent in these structures,
\cite{PhysRevLett.67.3598,CELINSKI1991L25} 
the formation of QWS are well confirmed. Indeed, \num{10} years later, 
MAE oscillations 
were observed in Cu(001)/Co, Ag(001)/Fe\cite{doi:10.1063/1.3670498} 
and Fe/Cu, Co/Cu structures,\cite{PhysRevB.87.134401} 
and the origin of these oscillations were attributed to QWS. 
Also, QWS induced oscillatory IEC was found in Co/MgO/Co PMA MTJ.\cite{PhysRevB.81.220407} 
Recent first-principles studies have correlated QWS with MCAE in Ag/Fe 
and IEC in Fe/Ag/Fe structures. \cite{Chang2015,Chang2017}
These works indicate that the influence of QWS on magnetic properties, 
specifically MAE, may become salient in some structures.

In this paper, we report \textit{ab-initio} calculations of MCAE in 
CFA/Ta structures and observe MCAE oscillations associated with the 
Ta layer thickness. These oscillations are further proved as induced by 
both majority-spin and minority-spin QWS confined in Ta layers. 
The origin of the significant 
MCAE oscillation is attributed to the repeated traversing of QWS across 
Fermi energy and the large SOC of Ta. 
In all, QWS formed in the capping layer provide us a unique 
method to tune MAE in the MTJ structure.


\section{methods}

Calculations were performed using Vienna \textit{ab initio} simulation 
package (VASP) based on projector-augmented wave (PAW) method and a 
plane wave basis set.\cite{PhysRevB.54.11169} The exchange and 
correlation terms were described using generalized gradient 
approximation (GGA) in the scheme of Perdew-Burke-Ernzerhof (PBE) 
parameterization.\cite{PhysRevB.59.1758} We used a kinetic energy 
cutoff of \SI{520}{\electronvolt} and a Gamma centered Monkhorst-Pack 
$k$-point mesh of \num{25 x 25 x 1}. The convergence of MCAE 
relative to $k$-point has been checked carefully, the variation 
of MCAE is about \SI{0.05}{\milli\electronvolt} when changing 
$k$-point mesh from \num{20 x 20 x 1} to \num{25 x 25 x 1}, 
which is at least a magnitude smaller than the oscillation amplitude 
of MCAE. The energy convergence criteria of all the calculations were 
set as \SI{1.0e-7}{\electronvolt}, and all the structures were relaxed 
until the force acting on each atom was less than 
\SI{0.01}{\electronvolt/\angstrom}. 
All the structures have at least \SI{15}{\angstrom} vacuum space 
to eliminate interactions between periodic images.

\begin{figure}[htb]
\includegraphics[width=0.48\textwidth]{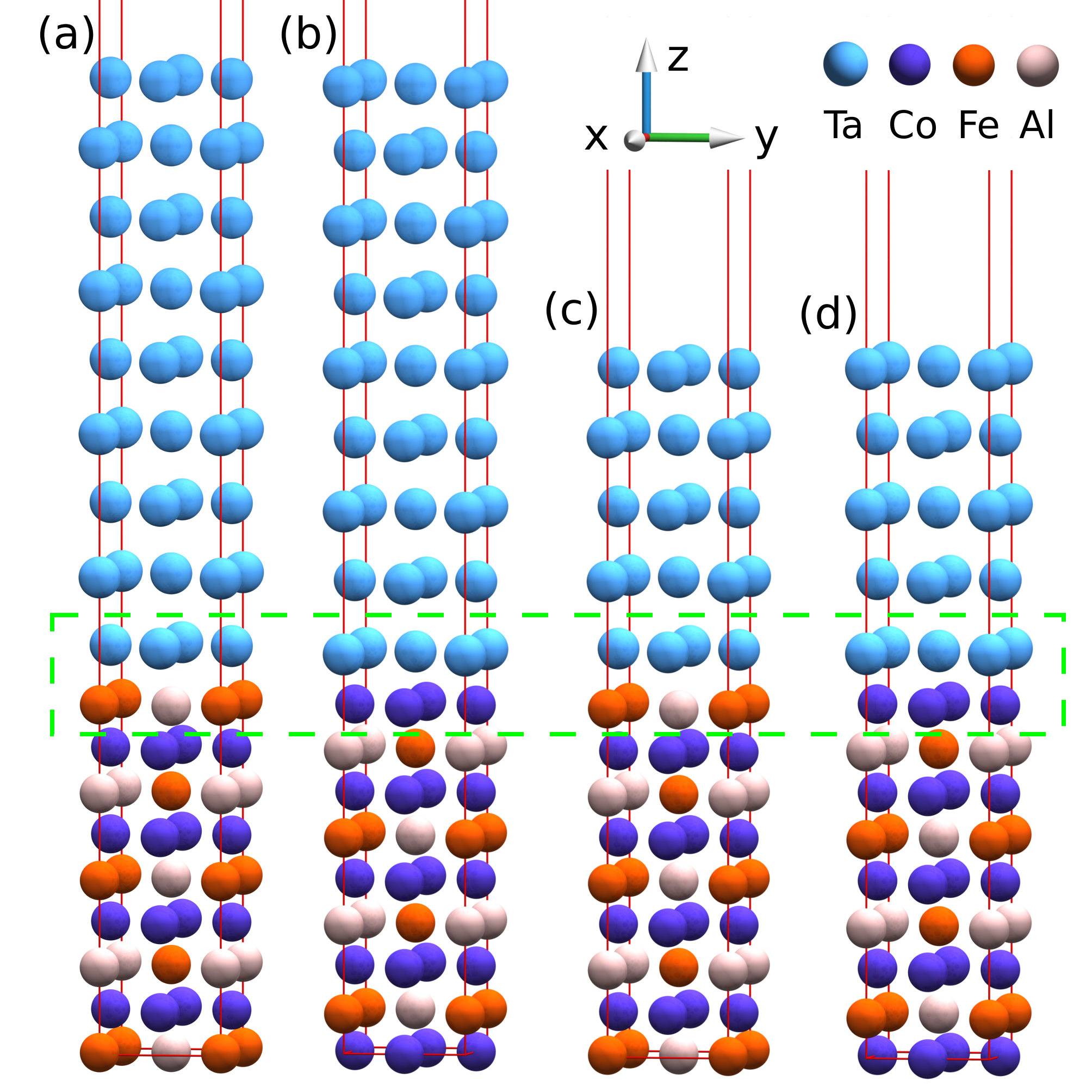}
\caption{\label{fig:struct} Crystal structure of (a)FeAl-CFA/Ta[\num{9}], 
(b)Co-CFA/Ta[\num{9}], (c)FeAl-CFA/Ta[\num{5}] and (d)Co-CFA/Ta[\num{5}]. 
Only 4 structures 
are shown as illustrations. In other structures, only the numbers of 
Ta ML are varied, ranging from \numrange{1}{12}. The dashed green 
rectangle box highlights the area of the interfaces.}
\end{figure}

Bulk CFA has a cubic $L2_1$ crystal structure. After fully relaxing 
bulk structure in volume and shape, the lattice constant is found to be 
$a_{bulk} = \SI{5.70}{\angstrom}$, 
perfectly matches the experimental value \SI{5.73}{\angstrom}. 
\cite{doi:10.1063/1.3549581} 
For CFA/Ta heterostructure, an in-plane lattice constant of 
$a = a_{bulk} / \sqrt{2} = \SI{4.03}{\angstrom}$ is adopted for 
the unit cell, which is rotated by \num{45} degrees 
from the conventional cell of bulk CFA. For all the CFA/Ta structures, 
\num{9} monolayers (ML) of CFA are used, and \num{1} to \num{12} ML of 
Ta layers are put on top of CFA, 
as shown in Fig. \ref{fig:struct}. 
We use CFA/Ta[$n$] to label structures of different Ta ML, 
where $n$ is the number of Ta ML, ranging from \num{1} to \num{12}. 
As for the interface between CFA and Ta, there exist two kinds of 
configurations and both of them have been investigated. 
FeAl-CFA/Ta is used as the label when FeAl layer of CFA 
directly contact with Ta, while Co-CFA/Ta is used when Co layer of 
CFA contact with Ta. 

To calculate MCAE, two-step procedures were adopted. Firstly, charge 
density was acquired self-consistently without taking into account 
SOC. Secondly, reading the self-consistent charge density, two 
non-self-consistent calculations were performed including SOC, 
with magnetization pointing towards the $[100]$ direction and
the $[001]$ direction, respectively. Finally, MCAE was calculated by 
$MCAE = E^{[100]} - E^{[001]}$, positive MCAE stands for PMA 
while negative MCAE for in-plane magnetic anisotropy.

To get a deeper understanding of the origin of oscillation, MCAE is 
decomposed into $k$-space. According to force theorem,
\cite{PhysRevB.47.14932,PhysRevB.50.9989,WANG1996337} 
the main contribution of MCAE originates from the difference of 
eigenvalues between two magnetization directions. Indeed, we found 
that the ion Ewald summation energy, Hartree energy, exchange 
correlation energy and external potential energy are exactly the 
same between two magnetization directions, the difference of total 
energy only comes from the difference of eigenvalue summation, this 
testifies the feasibility of $k$-space decomposition of MCAE. 
Specifically, this can be expressed as

\begin{equation}
MCAE(\bm{k}) = \sum_{i} n_{i,\bm{k}}^{[100]} \epsilon_{i,\bm{k}}^{[100]}
 - \sum_{i^\prime} n_{i^\prime,\bm{k}}^{[001]} \epsilon_{i^\prime,\bm{k}}^{[001]} 
 \label{equ:mcae_k}
\end{equation}

where $\bm{k}$ is the $k$-point index, $i, i^\prime$ are the band indexes of 
magnetization direction along [100] and [001], respectively. $n_{i,\bm{k}}$ is 
occupation number of this band, $\epsilon_{i,\bm{k}}$ is the energy of band $i$ 
at $k$-point $\bm{k}$.

\section{results and discussion}

\subsection{MCAE oscillation}

\begin{figure}[htb]
\includegraphics[width=0.48\textwidth]{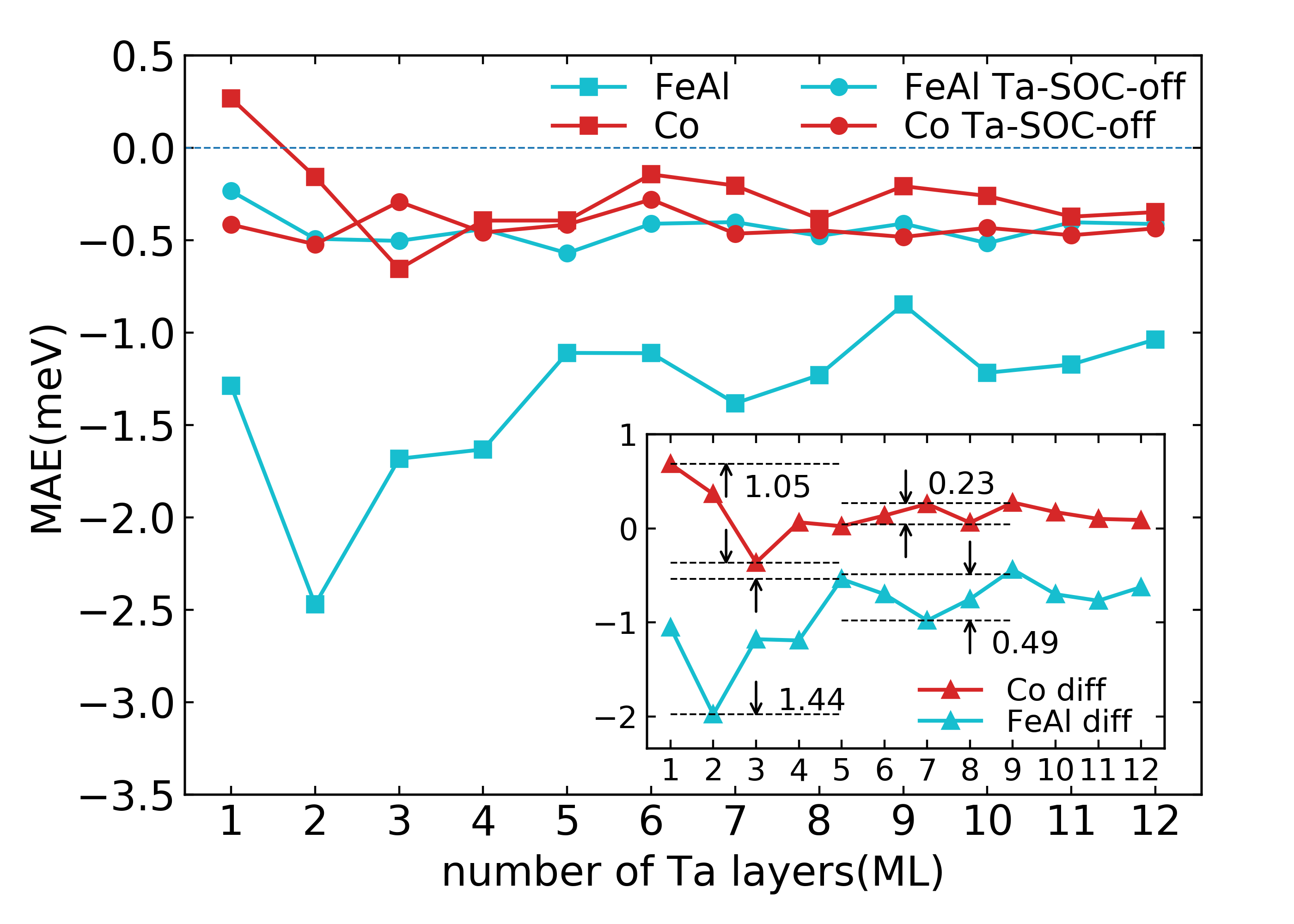}
\caption{\label{fig:oscil} MCAE oscillation with respect to Ta ML number. 
Cyan color for FeAl-interface structure, red for Co-interface 
structure. Circle mark for MCAE calculation with SOC of Ta switched 
off while SOC of CFA still included, square mark for normal MCAE 
calculation. Lines in the inset are defined by Eq. (\ref{equ:mcae_diff}). 
The numbers in the inset mark the oscillation magnitude of the 
first period and the second period.}
\end{figure}

Unlike the FM/oxide structure where the MCAE can be accounted as local 
hybridization of the interfacial Fe-$3d$ orbital and the interfacial O-$2p$ 
orbital,\cite{RevModPhys.89.025008} the MCAE of CFA/Ta structure varies 
strongly when the Ta thickness changes. In this circumstance, MCAE cannot 
be treated as a local property of the interface. We observe 
a strong oscillation of MCAE in FeAl-CFA/Ta structure relative to 
the thickness of capping layer Ta, as shown in Fig. \ref{fig:oscil}. 
The oscillation period is approximately \num{4} ML, and the oscillation 
amplitude decreases as the number of Ta ML increases. This is due to that 
the confinement effect of QW will become less prominent when the width of 
QW increases and the bulk states of Ta will 
account for a larger proportion in all the electron states. 
Interestingly, the oscillation is smaller in the Co-CFA/Ta 
structure, the reason for this phenomenon will be discussed later.

To comprehend the origin of the oscillations, we manually tweak the  
strength of SOC in the structures. Since MCAE only comes from SOC, 
switching off the SOC of Ta will totally screen out the contribution of Ta 
to MCAE of the whole system. For the FeAl-CFA/Ta structure, by 
suppressing the SOC of Ta while still keeping the SOC of CFA, 
oscillation of the MCAE relative to Ta layer thickness disappears 
[see cyan lines in Fig. \ref{fig:oscil}]. For the Co-CFA/Ta structure, 
a smaller oscillation exists and the tweaking of the 
SOC of Ta has little influence on the MCAE [see red lines in Fig. \ref{fig:oscil}]. 
These strongly indicate that the electron states in Ta play the 
determinant role in the MCAE oscillations of CFA/Ta structures.

A further analysis can be carried out by defining the MCAE difference 

\mathchardef\mhyphen="2D 
\begin{equation}
MCAE^{diff}(n) = MCAE(n) - MCAE^{Ta\mhyphen off}(n)
 \label{equ:mcae_diff}
\end{equation}

where $n$ is the number of Ta ML, $MCAE^{Ta\mhyphen off}(n)$ is the result 
calculated with SOC of Ta switched off. The $MCAE^{diff}(n)$ will only contain 
MCAE contribution originated from Ta layers, as plotted in the inset 
of Fig. \ref{fig:oscil}. Note three major differences can be discerned. 
Firstly, the oscillation magnitude of Co-CFA/Ta 
vanishes much faster than that of FeAl-CFA/Ta. Secondly, 
we can define an oscillation period of \num{4} ML in FeAl-CFA/Ta 
but it is harder to clearly define an oscillation period for 
Co-CFA/Ta. Thirdly, in Co-CFA/Ta, the mean value of $MCAE^{diff}(n)$ 
is essentially zero while the mean value of $MCAE^{diff}(n)$ of 
FeAl-CFA/Ta largely deviates from zero. The oscillation of physical 
properties relative to film thickness is a hallmark of QWS, and 
these three remarkable differences suggest that for FeAl-CFA/Ta, 
the electron states in Ta layers 
may form QWS and explain the MCAE oscillation. While for Co-CFA/Ta, 
since the MCAE oscillation is less prominent, there is less probability 
to correlate MCAE oscillation with QWS in Ta layers. The subsequent 
paragraph will concentrate on the analysis of MCAE with special electron states 
and evidence of the existence of QWS in FeAl-CFA/Ta structure. 
The same analytic procedures are also applied to Co-CFA/Ta in the 
Supplemental Material.

\subsection{Critical $k$-points and band structure}

\begin{figure}[htb]
\includegraphics[width=0.48\textwidth]{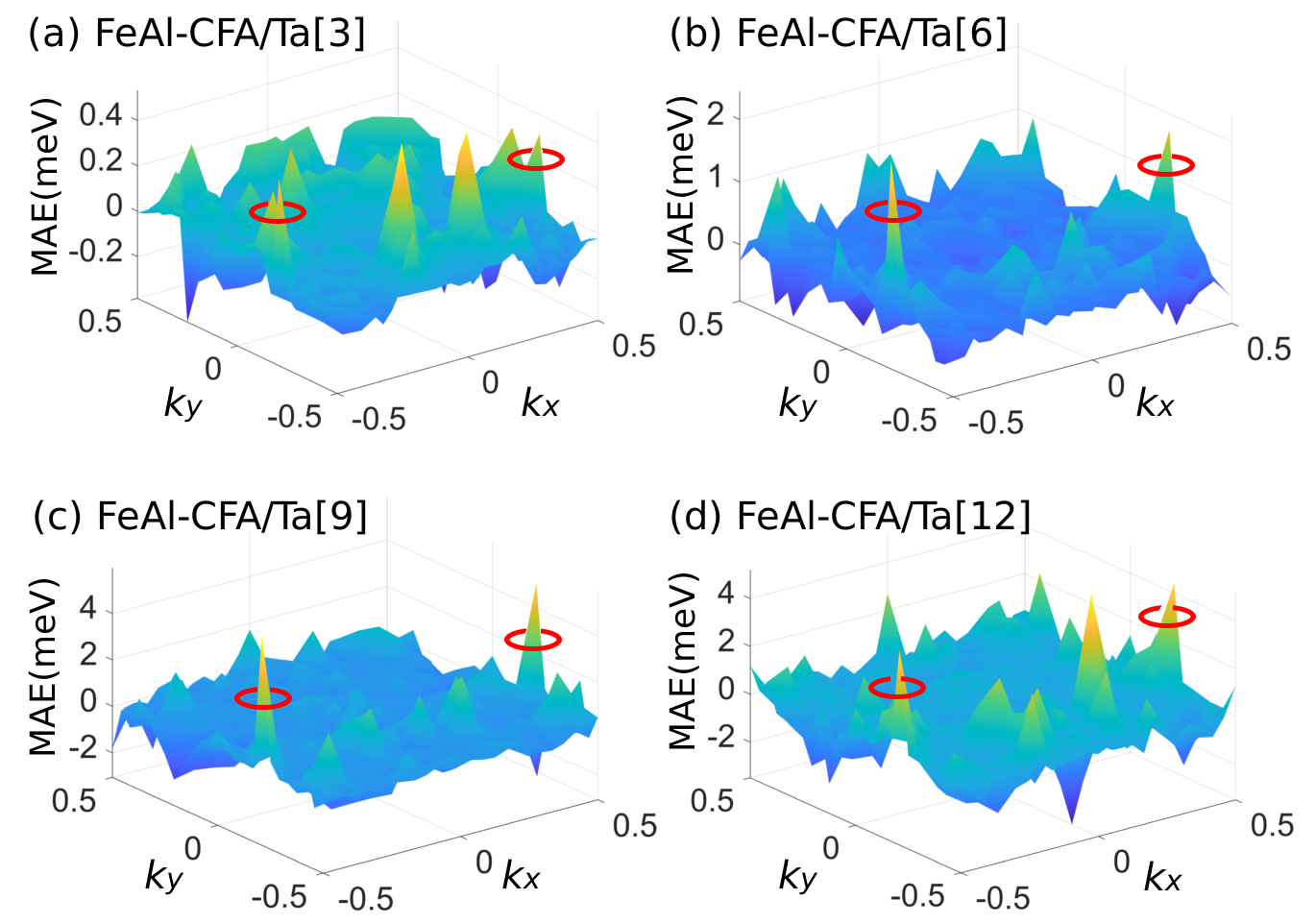}
\caption{\label{fig:kgraph} $k$-resolved graphs of MCAE in 2D-BZ for 
(a) FeAl-CFA/Ta[\num{3}], (b) FeAl-CFA/Ta[\num{6}], (c) FeAl-CFA/Ta[\num{9}] and 
(d) FeAl-CFA/Ta[\num{12}]. Critical $k$-points at 
$[k_x, k_y] = [-0.48, -0.2]$ and $[k_x, k_y] = [0.48, -0.2]$ are 
labeled by red circles, which have the dominant contributions to 
total MCAE.}
\end{figure}

Employing $k$-space resolved method, we dissect MCAE into 
two-dimensional Brillouin zone (2D-BZ) for FeAl-CFA/Ta[$n$]. 
Comparing $k$-resolved 
graphs for structures with different Ta layer thickness, two critical 
$k$-points, which have large contributions to MCAE, can be selected out, 
i.e. $k$-points at 
$[k_x, k_y] = [-0.48, -0.2]$ and $[k_x, k_y] = [0.48, -0.2]$, as 
shown in Fig. \ref{fig:kgraph}. 
(with number of $k$-points set as \num{25 x 25 x 1}, $k_x$ and 
$k_y$ ranging from $-0.48$ to $0.48$). The SOC breaks the symmetry 
of 2D-BZ, contributions to total MCAE slightly differ from each 
other between these two $k$-points. In fact, when considering 
the symmetry of the crystal structure, these two $k$-points 
are identical and locate at the center of $M$ point and $X$ point 
of the 2D-BZ, and they will be called as critical $k$-point 
in the following text. 

According to second order perturbation theory, the perturbation of 
SOC to one-electron energies can be written as 

\begin{equation}
\delta \epsilon_{i} = \sum_{i^{\prime} \neq i} 
  \frac{|\langle i^\prime | H_{SOC} | i \rangle |^2}
       {\epsilon_i - \epsilon_{i^\prime}}
\label{equ:perturb_eigen}
\end{equation}

\begin{equation}
\begin{split}
E_{corr}^{axis} = \sum_{i} n_i \delta\epsilon_i 
= & \frac{1}{2} \sum_{i} \sum_{i^{\prime} \ne i} 
\frac{n_i - n_{i^\prime}}{\epsilon_i - \epsilon_{i^\prime}} 
| \langle i^\prime | H_{SOC}^{axis} | i \rangle |^2, \\
& axis = [100], [001]
\label{equ:perturb_mcae}
\end{split}
\end{equation}

where $i, i^\prime$ is the quantum number of one-electron states, 
$\epsilon_i$ is the one-electron energy and $H_{SOC}$ is the 
Hamiltonian of SOC, $axis$ is the magnetization direction, 
and $E_{corr}^{axis}$ is the energy 
correction to unperturbed state caused by SOC.

\begin{figure}[htb]
\includegraphics[width=0.48\textwidth]{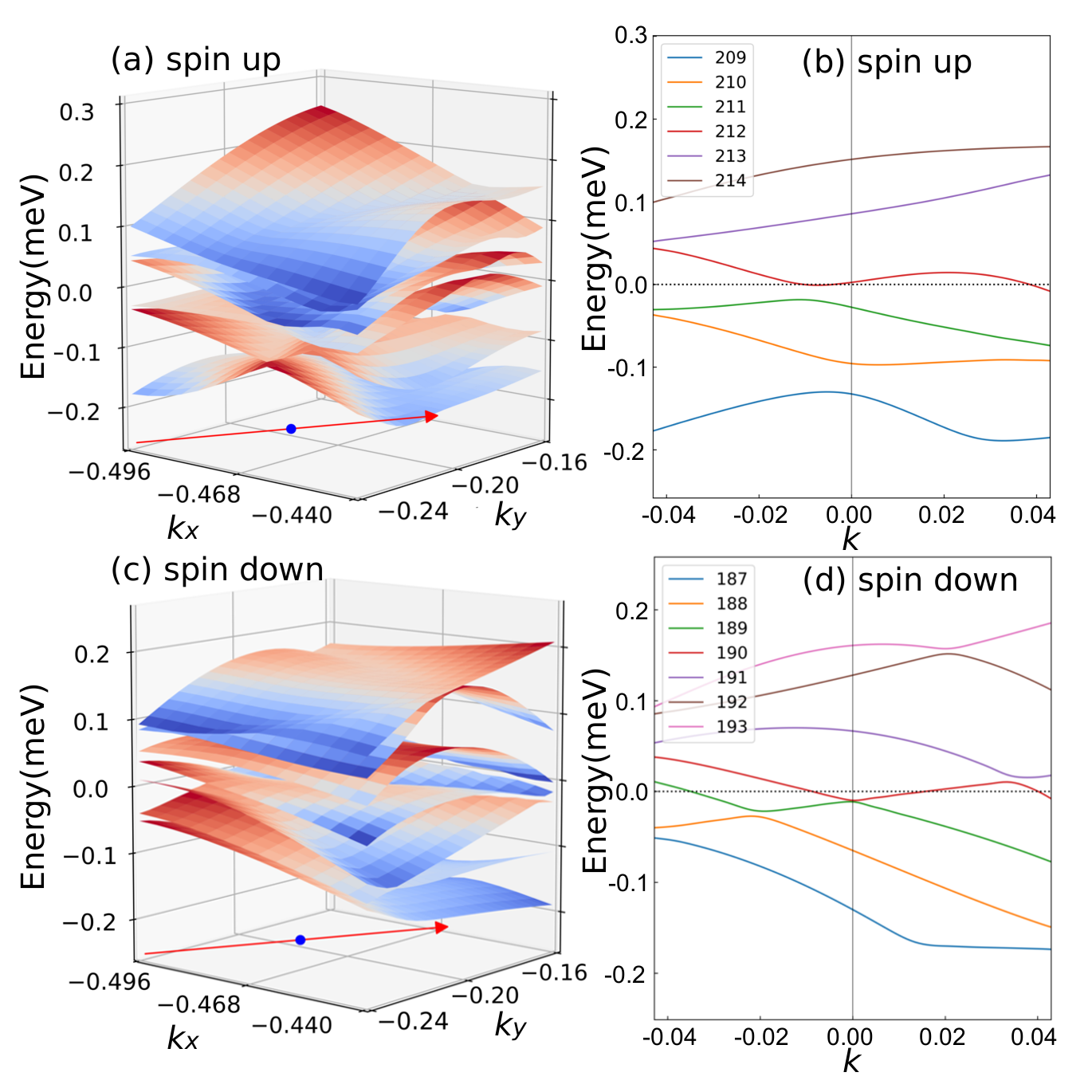}
\caption{\label{fig:band3d} (a) 3D spin-up band structure of FeAl-CFA/Ta[\num{9}] 
in a rectangle region around the critical $k$-point 
$[k_x, k_y] = [-0.48, -0.2]$, which is marked as a blue dot 
in the horizontal plane. Color represents the relative magnitude 
of energy in the same band, larger value in red, smaller value 
in blue. (b) Spin-up band structure along line: $k_y = 2.57 k_x + 1.03$ 
as an example, i.e., the red arrowed line in (a). 
(c) 3D spin-down band structure of FeAl-CFA/Ta[\num{9}] 
in the rectangle region. 
(d) Spin-down band structure along line: $k_y = 2.57 k_x + 1.03$.
The numbers in legend of (b) and (d) are the band indexes. 
These bands are vicinal 
to Fermi energy and have large contributions to total MCAE.}
\end{figure}

This expression indicates that 
only electron states near 
Fermi energy have maximal impact on MCAE. To extract out more 
information about states contributing most to the MCAE oscillation, 
we systematically examine spin-resolved band structures along different 
directions at this $k$-point. As an example, we draw the 
band structure along line $k_y = 2.57 k_x + 1.03$ , with band 
index ranging from \numrange{209}{214}, as shown in 
Fig. \ref{fig:band3d}(b) for spin-up electrons, and band 
\numrange{187}{193} for spin-down electrons, as shown in Fig. \ref{fig:band3d}(d).  
We find that the spin-up band with index \num{212} and spin-down band 
with index \num{190}
traverse Fermi energy along most of the directions 
in 2D-BZ. As a reminder, this particular number has no physical 
meaning but only labels the order of Kohn-Sham eigenvalues in 
the calculation result. Note due to exchange splitting, the energy of 
spin-down band are higher than the corresponding spin-up band which 
has identical band index with the spin-up band, so the Fermi-energy-vicinal 
bands are different for spin-up and spin-down electrons and should be 
considered separately.

For the Co-CFA/Ta, we find that different from FeAl-CFA/Ta, 
the rapid variations in 2D-BZ make the ascription of MCAE oscillation to a 
specific electron state a bit more difficult. We can still select out 
critical $k$-points but the magnitude of the peak does not have a 
sharp contrast with other $k$-points, as shown in Fig. \ref{sm:fig:kgraph} 
of Supplemental Material.

\subsection{Characterization of QWS}

\begin{figure}[htb]
\includegraphics[width=0.48\textwidth]{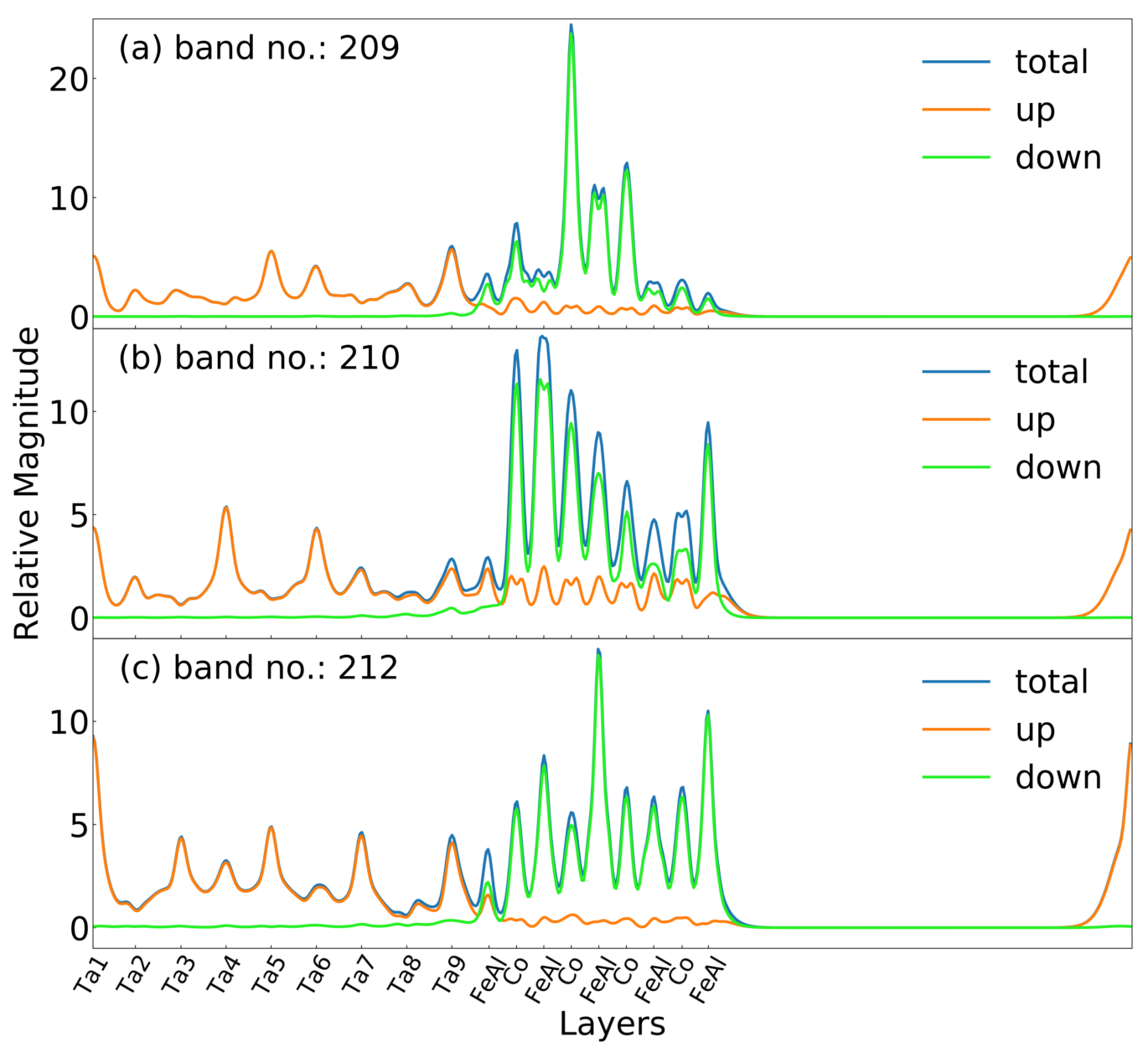}
\caption{\label{fig:parchg_up} Charge densities of energy bands at index 
(a) \num{209}, (b) \num{210} and (c) \num{212}. Green color for 
spin-down electron, orange for spin-up electron and blue for total 
charge density. The horizontal axes of these figures correspond to 
$z$ axis of the crystal structure. Note the energies correspond to 
the spin-up electrons of these bands are vicinal to 
Fermi energy, while energies correspond to the spin-down electrons 
are higher than Fermi energy due to exchange splitting. The spin-up 
electrons are mostly confined in Ta layers.}
\end{figure}

\begin{figure}[htb]
\includegraphics[width=0.48\textwidth]{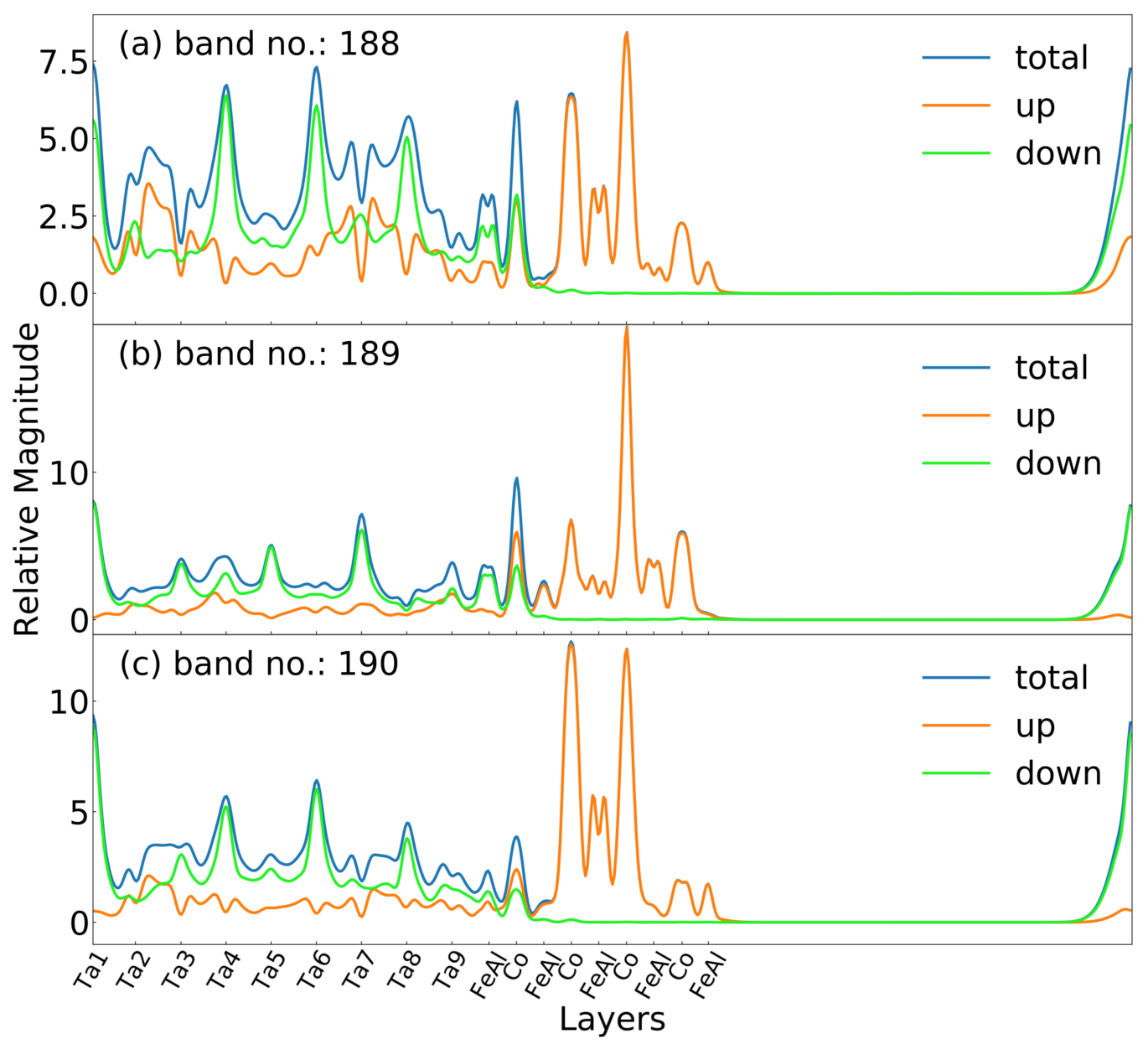}
\caption{\label{fig:parchg_dn} Charge densities of energy bands at index 
(a) \num{188}, (b) \num{189} and (c) \num{190}. 
Note the energies correspond to 
the spin-down electrons of these bands are vicinal to 
Fermi energy. The spin-down electrons of these bands are confined in 
Ta layers, while spin-up electrons couple to the states in CFA to some
extent.}
\end{figure}

To explore the nature of these specific electron states, the band-decomposed 
charge densities of these Fermi-energy-vicinal states are plotted, 
and we conclude that these are the quantum well states confined 
between the FeAl-CFA/Ta interface and the Ta/vacuum surface, as shown in 
Fig. \ref{fig:parchg_up} for spin-up electrons and 
Fig. \ref{fig:parchg_dn} for spin-down electrons. Note spin-up electrons 
of band \numrange{209}{212} are Fermi-energy-vicinal, while for spin-down 
Fermi-energy-vicinal states, their band indexes are from 
\numrange{188}{190}. For the Fermi-energy-vicinal states of both 
majority spin and minority spin, i.e. spin-up and spin-down electrons, 
all of them are mostly confined in Ta layers, as indicated by the 
orange lines in Fig. \ref{fig:parchg_up} and green lines in 
Fig. \ref{fig:parchg_dn}. 
With increasing band index, more 
wave crests are formed, which are the characteristic feature of QWS. 
The energies of QWS depend on the width of the well, namely, the 
thickness of the Ta layer. By increasing or decreasing Ta layer 
thickness, the QWS will fall or rise through Fermi energy, 
consequently lead to the oscillation of the total MCAE, as suggested 
by Eq. (\ref{equ:perturb_mcae}).

For the Co-CFA/Ta structure, the band-decomposed charge densities of 
Fermi-energy-vicinal states do not perfectly resemble that of an idealized 
one-dimensional quantum well with infinite potential barriers, 
the character of QWS is less apparent 
than FeAl-CFA/Ta. Band-decomposed charge densities of Co-CFA/Ta[9] are 
plotted in Fig. \ref{sm:fig:parchg_up} and Fig. \ref{sm:fig:parchg_dn} 
of Supplemental Material.

Before concluding the charge density analysis, we would like to clarify 
that both Fig. \ref{fig:parchg_up} and Fig. \ref{fig:parchg_dn} are 
band-decomposed, while the total charge density should be the sum of 
all the occupied bands. So the crests in the band-decomposed charge 
densities do not imply a large antiferromagnetic coupling between 
Ta layers and CFA. The magnetic moments of all the atoms are plotted in 
Fig. \ref{sm:fig:magmom_feal} for FeAl-CFA/Ta[9] and \ref{sm:fig:magmom_co} 
for Co-CFA/Ta[9] for the interested reader. 

\subsection{Interface potential}

\begin{figure}[htb]
\includegraphics[width=0.48\textwidth]{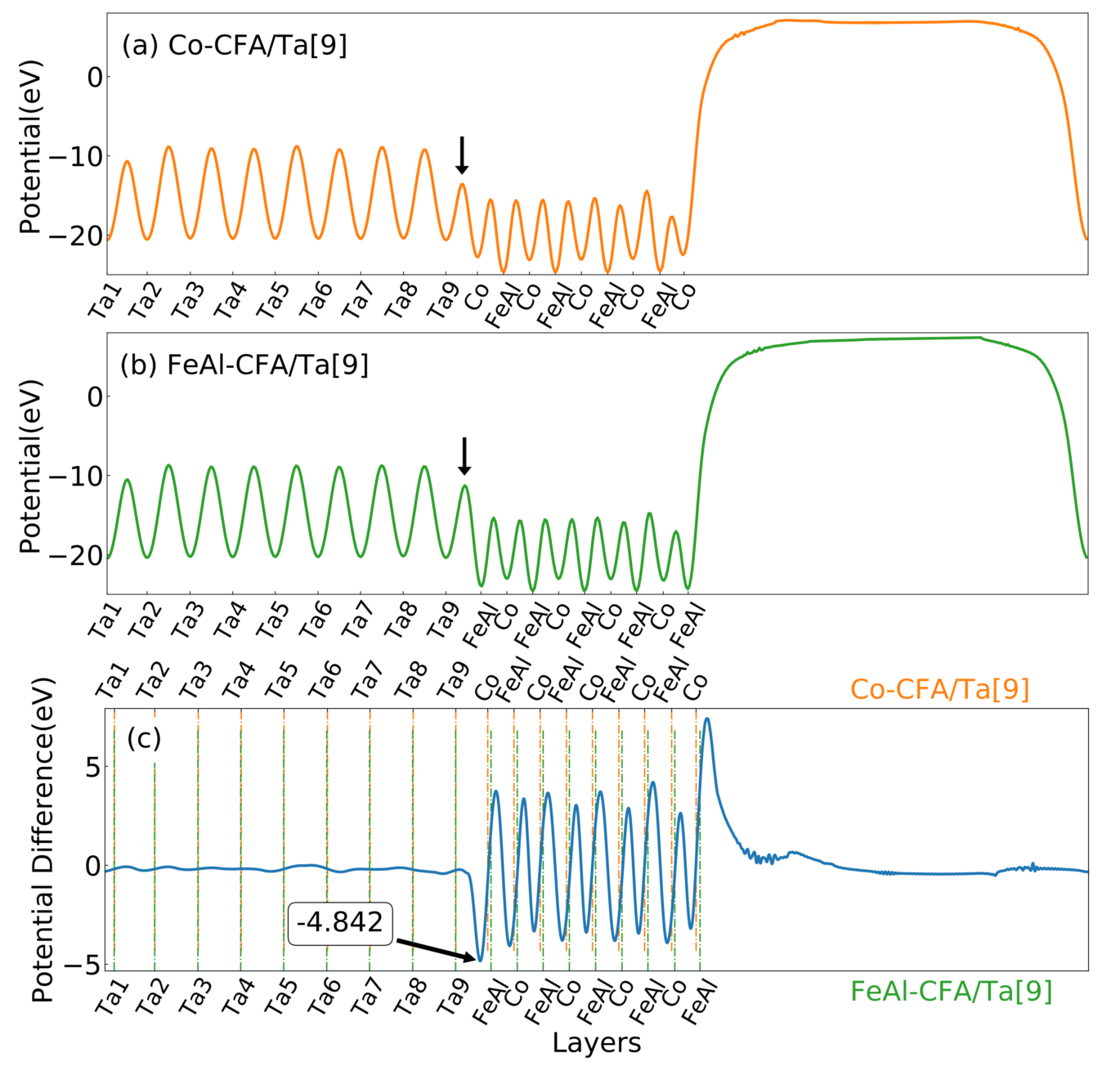}
\caption{\label{fig:pot} Potentials along crystallographic 
$z$ direction of (a) Co-CFA/Ta[\num{9}] and (b) FeAl-CFA/Ta[\num{9}]. 
(c) Potential difference between Co-CFA/Ta[\num{9}] structure and 
FeAl-CFA/Ta[\num{9}] structure, where the upper horizontal axis is 
for Co-CFA/Ta[\num{9}] and the lower horizontal axis is for 
FeAl-CFA/Ta[\num{9}]. The black arrows in (a) and (b) mark the 
location of the interfaces. The dashed orange and green lines represent 
$z$ coordinates of each layer in Co-CFA/Ta[\num{9}] 
and FeAl-CFA/Ta[\num{9}], respectively. A potential difference of 
\SI{4.842}{\electronvolt} can be found between two structures 
at the interfaces.}
\end{figure}

The phenomenon that only the FeAl-CFA/Ta structure has strong MCAE oscillation 
can be understood by the interface potential difference. The potential 
drop at the interface determines the magnitude of confinement of 
electrons. Potential of FeAl layer differs substantially from Co 
layer. As shown in Fig. \ref{fig:pot}, a larger mismatch of the potential 
between Ta and FeAl-CFA is found while the mismatch between Ta and 
Co-CFA is much smoother. Since the constructions of initial 
structures and the processes of atomic relaxations inevitably lead to 
a small displacement along $z$ axis between two structures of 
different interfaces, and the potential difference strongly relies 
on the origins of coordinates in two structures, we choose part of 
Ta layers as sampling and minimize the square error so as to 
accurately align these two structures, i.e. the potential difference is 
calculated as 

\begin{equation}
V_{diff}(z) = V_{Co}(z) - V_{FeAl}(z + \delta)
\end{equation}
where 
\begin{equation}
\delta = \arg\min_{\epsilon} \{ \int 
(V_{Co}^{Ta}(z) - V_{FeAl}^{Ta}(z + \epsilon))^2 dz \}
\end{equation}

where $z$ is the coordinate along $z$ axis, 
$\delta$ is the displacement of FeAl-interface structure that 
accurately aligns two structures, 
$V_{Co}$ and $V_{FeAl}$ are the potentials of Co-CFA/Ta[\num{9}] 
and FeAl-CFA/Ta[\num{9}], respectively.
$V_{Co}^{Ta}$ and $V_{FeAl}^{Ta}$ are the potentials of 
Ta layers of Co-CFA/Ta[\num{9}] and FeAl-CFA/Ta[\num{9}], respectively. 
$V_{diff}(z)$ is the potential difference plotted in 
Fig. \ref{fig:pot}(c).

By carefully aligning Ta layers of FeAl-interface and 
Co-interface structures to the same position, little difference 
is found in Ta part between both structures. But at the interface, 
the potential is \SI{4.8}{\electronvolt} higher in FeAl-CFA/Ta than 
Co-CFA/Ta [see Fig. \ref{fig:pot}(c)]. This larger mismatch ultimately 
makes confinement effect more prominent in the FeAl-CFA/Ta structure, 
thus explaining the larger 
magnitude of MCAE oscillation. The smoother transition of potential 
in Co-CFA/Ta does not strongly confine electrons into Ta layers, 
consequently MCAE oscillation relative to Ta layer 
thickness vanishes faster in the Co-interface structure.

\section{conclusions}
By means of first-principles calculations, we observed a significant 
oscillation of MCAE as a function of heavy metal layer thickness 
in the CFA/Ta structure with FeAl interface. 
The origin of this oscillation can be attributed to electron 
states confinement in Ta layers. Through $k$-space analysis, states 
having the largest contribution to MCAE can be traced down into special 
$k$-points located at the center of $X$ point and $M$ point in 2D-BZ. 
Moreover, it was unveiled that the Fermi-energy-vicinal states contribute 
the most to total MCAE. The wave crests and troughs appearing 
in these bands indicate that these are the quantum well states confined in Ta 
layers. 
The smaller oscillation magnitude in Co-interface structure can be 
explained by the smoother potential transition at the Co/Ta interface, 
which imposes less confinement on electrons in Ta layers. 
This work clarifies that 
due to the QWS in capping layer, MCAE in CFA/Ta cannot be accounted as 
a local property of the interface. Other than commonly used approach of 
inducing MAE by FM/MgO interface, we demonstrate that QWS formed in 
the capping layer provide a way to artificially control MAE in 
nanostructures, which could promote the development of STT-MRAM.

\begin{acknowledgments}

The authors would like to thank the supports by the projects from 
National Natural Science Foundation of China (No. 61571023, 61501013 
and 61627813), Beijing Municipal of Science and Technology 
(No. D15110300320000), and the International Collaboration Project
(No. 2015DFE12880 and No. B16001).


\end{acknowledgments}

%

\end{document}